\documentclass{mn2e}
\usepackage{mnras_cite}
\usepackage{psfig}

\def\gsim{~\rlap{$>$}{\lower 1.0ex\hbox{$\sim$}}}
\def\lsim{~\rlap{$<$}{\lower 1.0ex\hbox{$\sim$}}}

\def\d{{\rm d}}
\def\gife{GIF-{\sc ii}}

\begin{document}

\title[Orbital Parameters of Infalling Dark Matter Substructures]{Orbital Parameters of Infalling Dark Matter Substructures}
\author[A.~J.~Benson]{A.~J.~Benson\\ Department of Physics, University of Oxford, Keble Road, Oxford OX1 3RH, United Kingdom (e-mail: abenson@astro.ox.ac.uk)}

\maketitle

\begin{abstract}
We present distributions of the orbital parameters of dark matter
substructures at the time of merging into their host halo. Accurate
knowledge of the orbits of dark matter substructures is a crucial
input to studies which aim to assess the effects of the cluster
environment on galaxies, the heating of galaxy disks and many other
topics. Orbits are measured for satellites in a large number of N-body
simulations. We focus on the distribution of radial and tangential
velocities, but consider also distributions of orbital eccentricity
and semi-major axis. We show that the distribution of radial and
tangential velocities has a simple form and provide a fitting formula
for this distribution. We also search for possible correlations
between the infall directions of pairs of satellites, finding evidence
for positive correlation at small angular separations as expected if
some infall occurs along filaments. We also find (weak) evidence for
correlations between the direction of the infall and infall velocity
and the spin of the host halo.
\end{abstract}

\begin{keywords}
cosmology: theory - dark matter - galaxies: halos
\end{keywords}

\section{Introduction}

In currently favoured cosmological models, dark matter halos grow via
the merging together of smaller systems, leading to an ever-growing
hierarchy of halos. Recent numerical simulations have demonstrated
that the remnants of pre-existing dark matter halos which merged to
become part of a larger system (the ``host'') can survive for
significant periods of time within the larger system
\cite{moore99,klypin98}. These subhalos orbit around in the potential
of the host gradually losing mass via tidal forces and spiralling in
to ever smaller radii due to dynamical friction. These substructures
(or at least some subset of them) are presumably the abodes of
satellite galaxies, such as those found in the Local Group, and of the
majority of cluster galaxies.

This substructure has attracted a great deal of interest since its
discovery. Observational tests for its presence, though not yet
conclusive, are in good agreement with the theoretical expectations
\cite{metmad01,chiba01,dal01}. There has been much work conducted in
which the distribution and properties of substructures, their effects
on galaxy disks and so on were examined
\cite{ghigna98,tormen98,vdb99,zhang02,benson04,gao04,diemand04}. While
the orbital parameters of substructures have been measured in the past
this measurement has often been at the end point of the substructure
evolution (i.e. at the present day) when significant dynamical
evolution in the orbital parameters is expected
(e.g. \pcite{ghigna98}). Exceptions to this are the works of
\scite{tormen97}, \scite{vitvit02} and
\scite{khochfar04}. \scite{tormen97} and \scite{khochfar04} both
identified progenitors of halos in their N-body simulations and
measured the orbital parameters of them, while \scite{vitvit02}
searched for pairs of halos about to merge and measured the orbital
parameters of these. These works have typically made use of rather
small samples of orbits and (perhaps consequently) have been unable to
fully characterise the two dimensional distribution of orbital
parameters needed to construct realistic orbits.

The distribution of initial orbital parameters of substructure halos
at the time of merging into the host system is a particularly
interesting property as it represents the initial conditions which
determine the later evolution of the substructure within the host. The
effectiveness of many processes invoked to explain the morphological
transformation of galaxies in clusters (e.g. ram pressure stripping,
tidal mass loss, galaxy harassment, etc.) depend crucially on the
nature of the galaxy orbit (see, for example,
\pcite{moore98,abadi00}). The distribution of orbits will also
determine the rate of galaxy mergers and therefore the degree of
heating and rate of morphological transformation experienced by galaxy
disks. Taking a more practical point of view, recent semi-analytic
models of satellite halo orbits \cite{benson02,taybab04} have been
able to follow the orbital evolution of satellites quite accurately,
but these models are only as good as their initial conditions which,
until now, have been known only rather poorly.

In this work, we quantify the distribution of orbital parameters for
dark matter halos at the point of merging with their host (i.e. we
proceed in a similar way as did \pcite{vitvit02}). We measure this
distribution in a large number of N-body simulations to attain high
statistical precision and to facilitate checks of our techniques and
tests for variations of the distribution of orbital parameters with
variables such as redshift, halo mass etc. While we will present
distributions of orbital eccentricity and semi-major axis, our focus is
on distributions of radial and tangential velocities, which we find
are more practical when dealing with orbits in non-spherical systems
in which dynamical friction is at work\footnote{Since the orbital
parameters are constantly changing for such orbits, the eccentricity
and peri-centric distance no longer have the advantage of being
constant along the orbit. The orbital velocities are more closely
related to the quantities required by semi-analytic orbital models so
we prefer to use them. The two pairs of orbital parameters
(eccentricity+semi-major axis and radial+tangential velocity) are, of
course, equivalent.}. We also examine the distribution of infalling
substructures as a function of position on the virial sphere, and
explore correlations between orbital properties and the spin of the
host halo.

Our aim is to provide a precise and accurate measurement of the
distribution of orbital properties of substructures at the time of
merging, and to provide fits to this distribution so that it may be
used in further studies. This distribution could, in principle, depend
on many quantities, such as the masses of the merging halos, redshift,
cosmological parameters etc. Furthermore, the six parameters
describing each orbit (e.g. the position and velocity of the satellite
at the time of merging, or any equivalent parameter set) may well be
correlated with each other, such that we should really examine a
six-dimensional phase-space distribution function. With the currently
available N-body simulations we will limit ourselves to exploring a
two-dimensional function, typically that of radial and tangential
velocities (effectively assuming that infalling satellites are
uniformly distributed on a sphere around the halo centre and that
their tangential velocities have no preferred direction), although we
will explore correlations between these quantities and the host
halo. We note also that the situation could in principle be more
complicated still. We are aiming to quantify $P({\bf x})$, where ${\bf
x}$ are the orbital parameters and $P$ is the distribution of these
averaged over all merging events. However, after one merger with
parameters ${\bf x}_1$ the relevant distribution function for the next
merger may be different, $P({\bf x}|{\bf x}_1)$. An example might be
infall of halos along a filament. Knowing that one halo fell in from a
particular direction, it becomes more likely that the next halo will
fall in from a similar direction. We will explore one aspect of this
possibility by measuring the distribution of angles between pairs of
infalling satellites.

The remainder of this paper is arranged as follows. In
\S\ref{sec:analysis} we describe our analysis technique while in
\S\ref{sec:results} we present our results. We give our conclusions in
\S\ref{sec:discuss}.

\section{Analysis}
\label{sec:analysis}

\subsection{N-body Simulations}

To measure satellite orbital parameters we make use of a large number
of N-body simulations carried out by the VIRGO Consortium and which
are publicly available (see \pcite{jenkins98,kauffmann99,jenkins01}
for further details), together with one other simulation used for
testing various aspects of our methodology. These span a range of
cosmologies and redshifts. Details of the simulations used are given
in Table~\ref{tb:sims}. All of the outputs from these simulations are
analysed, but in practise only those at redshifts $z\lsim 2$ provide
statistically useful measurements of orbital parameter distributions.

\begin{table*}
\caption{The names, parameters and output redshifts of the N-body
simulations used in our analysis. The first two columns give the name
of the simulation set and the cosmological model respectively. Columns
3 lists the number of particles in each simulation, while columns 4
and 5 list the cosmological parameters $\Omega_0$ and $\Lambda_0$
appropriate to each simulation. Column 6 specifies the length of the
simulation cube, while column 7 specifies the mass of each particle in
the simulation. Column 8 gives the softening length used in the
simulation. Finally, column 9 lists the redshifts at which outputs
from the simulation are available.}
\label{tb:sims}
\begin{tabular}{ccccccccl}
\hline
Simulation & Model & Particles & $\Omega_0$ & $\Lambda_0$ & $L/h^{-1}$Mpc & $m_{\rm p}/h^{-1}M_\odot$ & $l_{\rm soft}h^{-1}$kpc & Redshifts \\
\hline
GIF & $\Lambda$CDM & $256^3$ & $0.3$ & $0.7$ & $141.3$ & $1.4 \times 10^{10}$ & 20 & 50, uniform in $\ln (a)$ from $z=50$ to $z=0$ \\
GIF & OCDM & $256^3$ & $0.3$ & $0.0$ & $141.3$ & $1.4 \times 10^{10}$ & 30 & 0.0, 0.1, 0.3, 0.5, 1.0, 1.5, 2.0, 3.0, 5.0 \\
GIF & SCDM & $256^3$ & $1.0$ & $0.0$ & $84.5$ & $1.0 \times 10^{10}$ & 36 & 0.0, 0.1, 0.3, 0.5, 1.0, 1.5, 2.0, 3.0, 5.0 \\
GIF & $\tau$CDM & $256^3$ & $1.0$ & $0.0$ & $84.5$ & $1.0 \times 10^{10}$ & 36 & 0.0, 0.1, 0.3, 0.5, 1.0, 1.5, 2.0, 3.0, 5.0 \\
\gife & $\tau$CDM & $256^3$ & $1.0$ & $0.0$ & $84.5$ & $1.0 \times 10^{10}$ & 36 & 0.0 \\
Virgo & $\Lambda$CDM & $256^3$ & $0.3$ & $0.7$ & $239.5$ & $6.86 \times 10^{10}$ & 25 & 0.0, 0.1, 0.3, 0.5, 1.0, 1.5, 2.0, 3.0, 5.0 \\
Virgo & OCDM & $256^3$ & $0.3$ & $0.0$ & $239.5$ & $6.85 \times 10^{10}$ & 30 & 0.0, 0.1, 0.3, 0.5, 1.0, 1.5, 2.0, 3.0, 5.0 \\
Virgo & SCDM & $256^3$ & $1.0$ & $0.0$ & $239.5$ & $2.27 \times 10^{11}$ & 36 & 0.0, 0.1, 0.3, 0.5, 1.0, 1.5, 2.0, 3.0, 5.0 \\
Virgo & $\tau$CDM & $256^3$ & $1.0$ & $0.0$ & $239.5$ & $2.27 \times 10^{11}$ & 36 & 0.0, 0.1, 0.3, 0.5, 1.0, 1.5, 2.0, 3.0, 5.0 \\ff
VLS & $\Lambda$CDM & $512^3$ & $0.3$ & $0.7$ & $479.0$ & $6.86 \times 10^{10}$ & 30 & 0.0, 0.5, 1.0, 2.0, 3.0, 5.0 \\
\hline
\end{tabular}
\end{table*}

\subsection{Group Finding}

In order to find merging dark matter halos in the simulations we must
first identify all dark matter halos. To locate dark matter halos in
the N-body simulations we employ two standard group finders, the
friends-of-friends (FOF; \pcite{defw}) and spherical overdensity (SO;
\pcite{lc94}) algorithms. We will compare results for halos found
using these two techniques to test for any dependence on the group
finding algorithm used.

Each algorithm has one tunable parameter, the linking length, $r_{\rm
link}$, for the FOF algorithm and the mean density contrast inside the
sphere, $\bar{\Delta}$, for the SO algorithm. Both can be related to
the mean density of dark matter halos (once a specific halo density
profile has been chosen in the case of the FOF algorithm). We apply
each algorithm twice, once assuming a mean overdensity for halos of
$\bar{\Delta}=18 \pi^2 \approx 177.7$ (equivalent to $r_{\rm link} =
0.20 \bar{r}$, assuming an isothermal halo profile\footnote{It is well
known that cold dark matter halos are not well approximated by
isothermal spheres. However, if we instead adopt an NFW density
profile \protect\cite{NFW} for our halos the appropriate value of
$r_{\rm link}$ ranges between $0.22\bar{r}$ and $0.26\bar{r}$ for
halos with concentrations in the range 5 to 15. As such, a somewhat
larger value of $r_{\rm link}$ may be appropriate. Nevertheless, we
will retain the convention of assuming isothermal halos here and
resign a study of the most appropriate linking length to use to future
work.}, where $\bar{r}$ is the mean inter-particle spacing in the
simulation), as expected from the spherical collapse model in a
critical density cosmology (e.g. \pcite{peebles80}), and once using
the mean overdensity expected from the spherical collapse model for
the specific cosmology and redshift in question \cite{lc93,eke96}. We
will refer to these two alternatives as ``fixed $\Delta$'' and
``variable $\Delta$'' respectively, and will compare results from the
two.

Once halos have been located by either algorithm we apply a procedure
to remove unbound halos from the resulting catalogue. Our technique is
described fully by \scite{benson01} and involves repeatedly removing
the least bound particle from an unbound halo until the halo either
becomes bound, or falls below the minimum mass required to be included
in our catalogue.

\subsection{Defining the Halo Centre}

To measure orbital properties of infalling satellites we need to
define a centre (both in position and velocity) for each halo in order
to have a suitable origin for our coordinate system. The simplest
option is to determine the centre of mass and the mass weighted mean
velocity of the halo and take these as the origin. We call this ``COM
centring''. Because of its simplicity we will examine results based
upon this approach. However, while a simple centre of mass estimate of
the halo centre is reasonable if halos are smooth, spherical systems,
in reality it has many failings (particularly when applied to FOF
halos). The FOF algorithm often links together halos that are about to
merge by a low density ``bridge'' of particles. This will skew the
centre of mass of the halo away from what perhaps should be considered
the centre (e.g. the position corresponding to the centre of mass of
the main component of the merging system). Because of this limitation
we will adopt a second approach in which we define the centre of a
halo as being the position of the particle with the lowest
gravitational energy (counting only interactions with other particles
in the halo). This will naturally pick out a particle in the densest
region and, given two halos joined by a low-density bridge should pick
out a particle in the more massive of the two halos. However, we
cannot take the velocity of this particle as being representative of
the velocity of the halo, since its motion will consist of the mean
halo motion plus a component due to the halo's internal velocity
dispersion. Unfortunately, just as in position space, halos in
velocity space can show bimodal distributions (as happens when a halo
is linked to a nearby halo which is infalling). This can bias the mass
weighted mean velocity estimate of the origin away from the
``correct'' value. To circumvent this problem we adopt a similar
approach in velocity space as in position space. Namely, we estimate
an ``energy'', $\epsilon$, for each particle, $i$, using $\epsilon_i =
\sum_{j\ne i} -1/|{\bf v}_i-{\bf v}_j|$, with the sum taken over all
particles in the halo, and then locate the particle with the lowest
energy. This should lie close to the true mean motion of the halo. We
call this method ``MBP centring''.

\begin{figure*}
\psfig{file=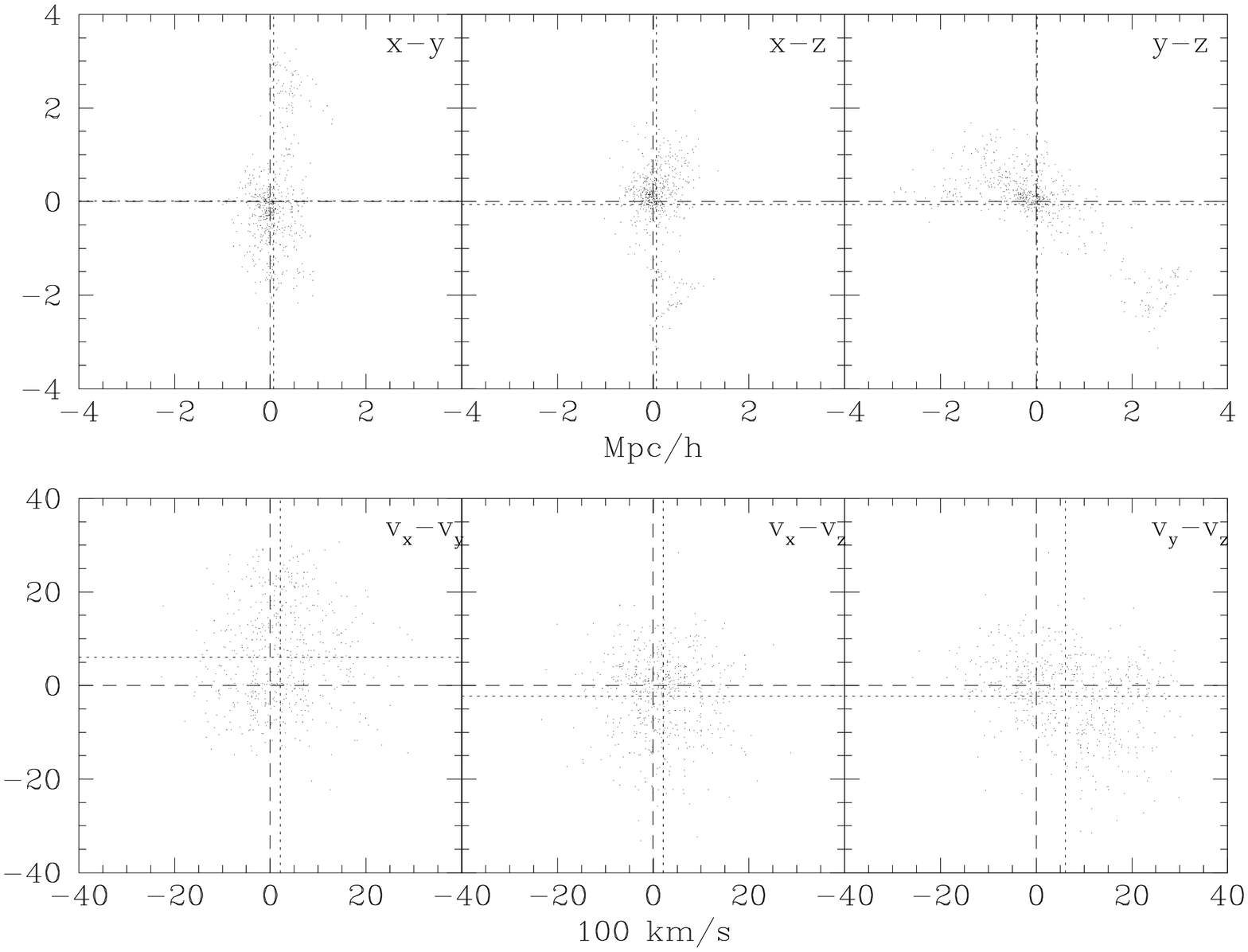,width=180mm,angle=270}
\caption{The results of using the COM and MBP algorithms to define the
origin of the coordinate system in a dark matter halo identified in
the $z=0$ output of the GIF $\Lambda$CDM simulation. The upper row
shows three projections of the spatial distribution of particles. The
intersection of the dashed lines indicates the origin according to the
MBP algorithm, while that of the dotted line indicates the origin
according to the COM algorithm. The lower row shows projections of the
same halo in velocity space. Dashed and dotted lines are as in the
upper row.}
\label{fig:halocentre}
\end{figure*}

It is worth noting that the velocity origin can differ significantly
between the two definitions we adopt. Fig.~\ref{fig:halocentre} shows
the centre of a particular halo from the $z=0$ output of the GIF
$\Lambda$CDM simulation in both position and velocity space. Each
frame has its origin on the most-bound particle, as marked by the
dashed lines, while the dotted lines indicate the centre of mass or
mass-weighted mean velocity. In this example, the centre of mass
almost coincides with the most bound particle (somewhat fortuitously
as nearby halos on either side, linked in by the FOF algorithm, are
cancelling each other out). In velocity space however, we see that the
velocity origin is shifted by over 500km/s from the more realistic
velocity origin. This could seriously affect our estimates of orbital
parameters.

\subsection{Satellite Orbital Parameters}

From our catalogue of dark matter halos in each simulation we search
for pairs of halos which are about to merge. From here on, all
velocities are measured in units of the virial circular velocity of
the host halo, $V_{\rm vir}$, and all radii in units of virial radius
of the host halo, $r_{\rm vir}$, as we expect these to be
characteristic scales of the systems\footnote{We convert the comoving
coordinates of the N-body simulation to physical coordinates and add
on the Hubble flow to the peculiar velocities taken from the N-body
simulations. Halo virial radii and velocities are determined from
their masses assuming the halo to have the mean density appropriate to
a just-collapsed spherical top-hat overdensity.}. We search for halos
within a distance from the host centre between $r=1\pm \Delta r$, and
which have an inward directed velocity, ${\bf v}$
(i.e. $\mathbf{r}\cdot\mathbf{v}<0$ where ${\bf r}$ is the vector from
the centre of the host to the centre of the potential satellite
halo). We choose $\Delta r = 0.2$. Note that we allow for the
possibility of halos with $r<1$ since the non-spherical shape of real
halos can permit a halo to remain separate from the host even when
$r<1$. It should be noted that this radial selection biases us against
finding mergers between halos of comparable mass (since in this case
it is unlikely that the satellite will remain as an isolated halo once
its centre is within $1+\Delta r$). For present purposes this bias is
unimportant, and so we retain the above criterion for simplicity. This
bias could however, be easily circumvented by adopting a radial
selection based upon the sum of the host and satellite virial radii
instead. From the halos selected in this way, we compute the radial
and tangential components of velocity. We also store the three
dimensional position and angular momentum of the merging satellite.

Since we are interested in the orbital parameters of satellites as
they cross the virial radius of a larger halo we correct our orbital
parameters (which are measured at some radius close to, but not equal
to, the virial radius). To do this we treat the two halos as point
masses, and simply determine the point at which the satellite's orbit
first crosses the virial radius of the larger halo. We store the
position, velocity and angular momentum of the satellite at this
point. This approach is an approximation for two reasons. Firstly, as
the host halo is not a point mass, the mass interior to the
substructure's orbit will change along that orbit. In practise this
effect is quite small, leading to only a 5\% error in the orbital
velocities. (Note also that the density profile is not spherically
symmetric, which will lead to further errors.) Secondly, we neglect
the effects of dynamical friction on the orbital parameters. A simple
estimate based upon Chandrasekhar's methods indicates that this would
lead to an error in our orbital velocities of around 10\% for a
substructure to host mass ratio of 0.08 (which is typical of the
systems found in our simulations), scaling approximately in proportion
to this ratio. All of these problems could be largely overcome by
solving the equations of motion for the substructure in a realistic
host potential including a dynamical friction term. This will be the
focus of future work.

Some fraction of substructures are found to be on unbound orbits. This
presents no problem for our analysisb, we can of course still measure
the orbital parameters of such substructures, and so we retain these
objects in our calculations. The fate of such substructures will be
discussed below. Some substructures are found with $r<1$---already
inside the virial radius by our definition. These substructures are
propagated backwards along their orbit to find their orbital
parameters at the time of crossing $r=1$ (as with all orbits, no
account is made for any mass loss which might have occurred from these
halos, nor for the effects of dynamical friction). Finally, we find
some halos whose orbits do not cross the virial radius of the
host. Such halos are flagged as being ``bad'' and are treated
separately from other halos (see \S\ref{sec:bad}).

We must also account for the fact that our selection of halos with
$1-\Delta r<r<1+\Delta r$ leads to a bias against finding radial
orbits as they will spend less time in this region than more circular
orbits. To correct for this we simply determine, from the measured
orbital parameters of the satellite, the time, $\delta t$, it takes to
traverse the region $r=1+\Delta r$ to $r_{\rm min}$. Here $r_{\rm
min}$ is the minimum radius at which the satellite halo would have
been identified by the group finder. When constructing distributions
of orbital parameters we then weight by $\Delta t/\delta t$ where
$\Delta t$ is the cosmic time between the current N-body simulation
output and the previous one (or $t=0$ in the case of the highest
redshift output).

The determination of $r_{\rm min}$ depends on the group finder
used. With the SO group finder it is relatively easy to determine
$r_{\rm min}$. Under the SO algorithm each halo is assigned a radius
(the radius containing a mean overdensity of some specified
value). Once all halos have been found any halo whose centre lies
within the radius, $r_{\rm SO}$, of a larger halo is merged with that
larger halo and removed from the list of individual halos. (Note that
the radius of the larger halo is not changed by this merging.) Thus,
$r_{\rm min}$ is simply $r_{\rm SO}$, or $1-\Delta r$, whichever is
larger.

For the FOF group finder things are a little more complicated as the
halos found are not spherical. The satellite halo would no longer have
been found as an isolated object by the group finding algorithm once
any one of its particles came within a distance $r_{\rm link}$ of a
particle in the larger halo. We therefore search for the first point
along the orbit of the satellite at which any one of its particles
comes within $r_{\rm link}$ of a particle in the larger halo. We
define $r_{\rm min}$ to be the orbital radius at this point, or
$1-\Delta r$, whichever is larger. The advantage of this approach is
that it works even for the non-spherical halos found by the FOF
algorithm. Its disadvantage is that it treats the orbit as that of two
point masses and also ignores any internal evolution of the satellite
or host halos during the time it takes the satellite to move along its
orbit. This latter is not a problem providing the two halos are in
internal equilibrium and not rotating since then, although the
individual particles in the halos move, their distribution at any time
provides a fair sample of the mass distribution of the halo at any
later time. Of course, in reality the halos will not be in equilibrium
(although we expect them to be close to it). In particular, the FOF
algorithm is known to make ``dumbbell-shaped'' halos by linking
together two halos by a low density bridge. These are certainly not
equilibrium systems in the sense used here. They are also those in
which the two-body orbit approximation is likely to be worst. We
consider this to be a limitation of the FOF algorithm, and do not
explore more complicated ways of dealing with this problem here.

It should be noted that, with our method for locating merging halos,
some host halos may be experiencing mergers with multiple
substructrues at any given time. In fact, we find that about 25\% of
all of our merger events at $z=0$ involve two or more substructures
accreting onto the same host halo. For the largest clusters we find up
to around twenty ongoing mergers in some cases. We find very few
mergers with low mass ratios (e.g. less than 4:1). As such, the
inclusion or not of hosts currently underdoing major mergers does not
affect our results significantly.

\section{Results}
\label{sec:results}

We examine the orbital parameter distributions for each individual
output of each simulation. We will also combine results together where
possible to improve the statistical precision. All results will make
use of the FOF halo finding algorithm, MBP halo centring and the
variable $\Delta$ method for setting $r_{\rm link}$/$\bar{\Delta}$
unless otherwise stated.

\begin{figure}
\psfig{file=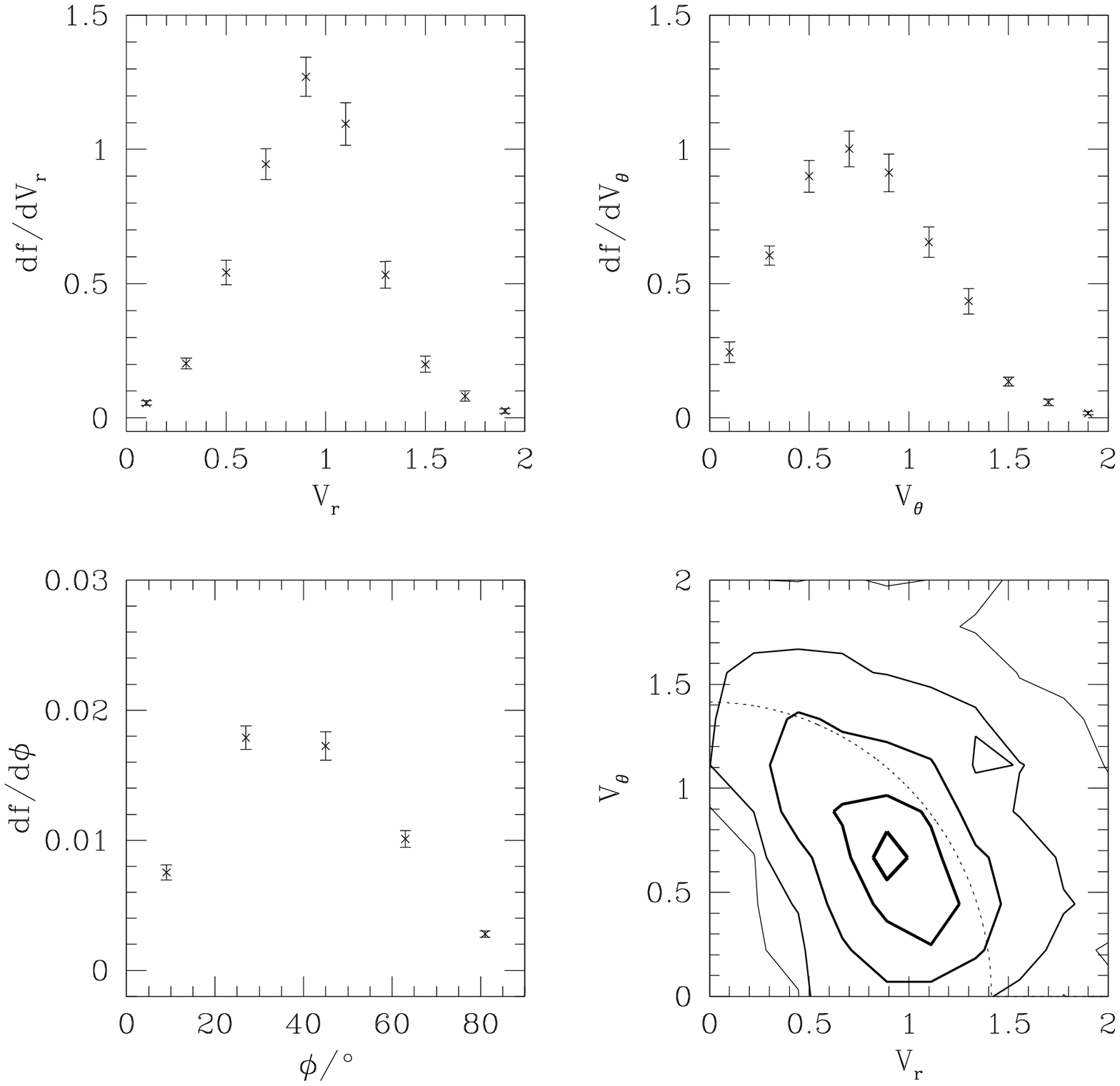,width=80mm}
\caption{Distributions of orbital parameters measured in the VLS plus
VIRGO $\Lambda$CDM $z=0$ output. Upper left and right-hand panels show
distributions of radial and tangential velocities respectively. The
lower left-hand panel shows the distribution of infall angles, while
the lower right-hand panel shows the two-dimensional distribution of
radial and tangential velocities. Contours are drawn at $\d^2 f/\d
V_{\rm r}\d V_\theta = 0.01$, 0.1, 0.5, 1.0 and 1.4 from lightest to
heaviest lines. The division between bound and unbound orbits in this
panel is shown by the dotted line.}
\label{fig:opexam}
\end{figure}

Figure~\ref{fig:opexam} shows an example of the distribution of
orbital parameters that we measure. The distribution of radial and
tangential velocities (upper left and right-hand panels respectively)
have quite simple, and perhaps unsurprising, forms, being peaked at
$V\sim 1$ with a dispersion of order unity. The infall angle,
defined as the (negative of the) angle between the infalling
substructure's radius and velocity vectors (i.e. $\phi = -\cos^{-1}
\mathbf{r}\cdot\mathbf{v}/|\mathbf{r}|/|\mathbf{v}|$), is shown in the
lower-left hand panel. This distribution will be investigated further
in \S\ref{sec:fitfunc}. Finally, the lower right-hand panel shows the
two-dimensional distribution of radial and tangential orbital
velocities. It is clear that there is a significant correlation
between these two parameters. Another interesting feature of this
distribution is that a significant fraction of orbits drawn from this
distribution are initially unbound. The energy of orbits, in our
units, is given by
\begin{equation}
E=-1+{1\over 2 f_2} \left( V_{\rm r}^2 +{(2-f_2)^2 \over f_2^2} V_{\theta}^2 \right),
\end{equation}
where $f_2=1+M_2/M_1$. Note that $f_2 \equiv M_2/\mu$ where $\mu=M_1
M_2/(M_1+M_2)$ is the usual reduced mass. The dotted line in
Fig.~\ref{fig:opexam} shows the line $E=0$ for the case $f_2=1$
(i.e. $M_1\ll M_2$). Points to the upper right of this line correspond
to unbound orbits. For the particular distribution shown about 18\% of
all orbits are unbound. We choose to retain these orbits for two
reasons:
\begin{enumerate}
\item When using the measured distribution to select initial orbits
for satellites, unbound orbits can easily be discarded if desired.
\item Due to the effects of dynamical friction, an orbit that starts
out unbound will not necessarily stay that way.
\end{enumerate}
To examine the importance of this second point we employ the
semi-analytic model of \scite{benson04} which follows the cosmological
growth of dark matter halos (and their associated galaxies) including
a detailed treatment of the orbital evolution of satellite halos. In
\scite{benson04} the initial orbits of merging satellites were
determined by setting the energy of each orbit equal to that of a
circular orbit at half the virial radius and choosing a circularity
(i.e. the angular momentum of the satellite in units of that of a
circular orbit with the same energy) from a uniform distribution
between 0.1 and 1.0. These choices were motivated by the results of
\scite{ghigna98}. Here, we instead use the measured distribution of
orbital velocities, including unbound orbits, to set the initial
velocity of satellites, and choose their initial position at random on
a sphere with radius equal to the virial radius of their host. From
this cosmologically representative sample of halos and orbits, we
identify those which start out unbound. Of these, some fraction will
lose sufficient energy through dynamical friction that they become
bound by the endpoint of their evolution (i.e. by $z=0$) while others
will fail to do so and will instead leave their host halo with
positive energy. We find that approximately 2\% of all initially
unbound orbits (equivalent to 0.3\% of all orbits) fail to become
bound and so escape their halo. As such, these ``lost'' satellites are
only a small fraction of the total.

\begin{figure}
\psfig{file=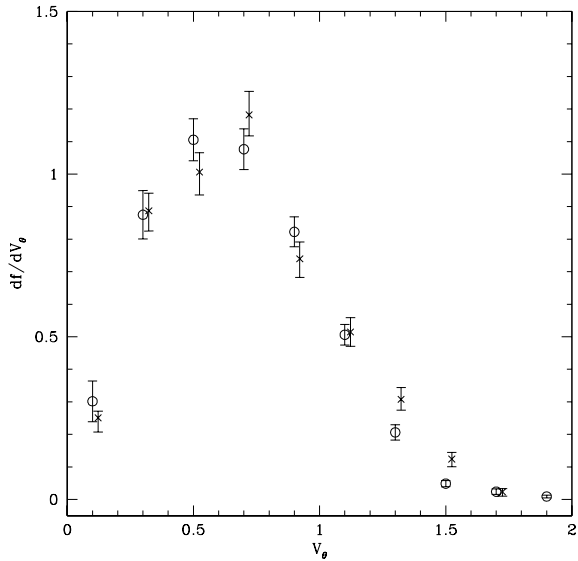,width=80mm}
\caption{The distribution of tangential velocities for orbits. Circles
show results for the VLS plus VIRGO $\Lambda$CDM simulations $z=0$
output from this work, while crosses (offset horizontally slightly for
clarity) show the results of \protect\scite{vitvit02}.}
\label{fig:vitvit}
\end{figure}

Our results are in good agreement with previous
work. Figure~\ref{fig:vitvit} shows a comparison of the distribution
of tangential velocities with that found by Vitvitska et al. (2002;
our $V_\theta$ is equivalent to their $L/L_{\rm vir}$). Although the
two distributions differ as judged by a $\chi^2$ test, the discrepancy
is due to two points and plausibly reflects our ignorance of the true
errors and the differences in the simulations (e.g. softening, method
of force calculation etc.) used in this work and that of
\scite{vitvit02}.

\subsection{Tests of the Distributions}

Firstly, we examine which, if any of our measured distributions are
consistent with each other. This will allow us to determine which
distributions we can realistically average together in order to
improve the statistical precision of our measurements.

\subsubsection{Calibration of $\chi^2$}

We adopt a simple $\chi^2$ test to determine if two of our measured
two-dimensional velocity distributions are consistent with each
other. It should be noted that the errors which we determine for our
distributions are likely to be an underestimate---they account for the
finite number of mergers in each bin, but ignore such contributions as
errors in our determinations of orbital velocities etc. Given this,
and the fact that our errors may not be normally distributed, we would
ideally like a calibration of the $\chi^2$ test. To achieve this we
compare distributions from our GIF and \gife\ $\tau$CDM $z=0$
simulations. Comparing both the FOF and SO results, with halo centres
defined using both centre of mass and most bound algorithms we find
values of $\chi^2$ per degree of freedom which scatter around unity,
with a mean of $1.05$. Although we would ideally like many more
independent simulations to test our errors this gives us confidence
that the errors are a good approximation to the true uncertainty on
each data point.

\subsubsection{Distribution With and Without ``Bad'' Orbits}
\label{sec:bad}

A small fraction of the orbits that we find are flagged as being
``bad'' in the sense that they do not pass through one or both of the
radial limits which we use for computing the weight to assign to each
orbit. This may represent cases in which a halo formed within the
outer radial limit (and so never passed through it), or, more likely,
a limitation of the simple, two-body orbit neglecting dynamical
friction that we use to approximate the motion of the halos. The best
guess at a suitable weight for these orbits is to use their
instantaneous radial velocity to determine the time taken to cross
between the two radial limits. However, we find that the resulting
distributions of orbital parameters for bad orbits differ
significantly (as judged by the $\chi^2$ test) from those of good
orbits. Therefore, we adopt the approach of excising all bad orbits
from our distributions. Ideally, we should deal with these better by
solving for the orbit correctly (i.e. including extended masses and
dynamical friction) to see if they really do merge and thereby
assigning a realistic weight.

\subsubsection{Number of particles per halo}

Our halo finding algorithms retain only halos consisting of ten
particles or more. To test whether particle number has any effect on
the measured distribution of orbit parameters we compare measurements
of the orbit distribution in the VLS and VIRGO simulations with the
equivalent GIF simulations, keeping halos with 10 or more particles in
the VIRGO and VLS simulations and adopting an equivalent mass cut in
the GIF simulations (49 or more particles per halo in the $\Lambda$CDM
and OCDM simulations and 227 or more particles in the SCDM and
$\tau$CDM simulations), such that the minimum mass of halos in each
simulation is the same (this avoids any consequences of possible
mass-dependent trends in the orbits).

We find no evidence of any significant difference between the velocity
distributions constructed from halos with 10 or more particles and
those with 5--20 times more particles from the GIF simulations. The
measured values of $\chi^2$ per degree of freedom are scattered around
unity and are consistent with being drawn from a $\chi^2$ distribution
(as judged by a K-S test).

While we would ideally like more extensive tests of the effects of
particle number\footnote{Ideally we would like a set of simulations
identical in all respects apart from the number of particles
used. This would permit direct comparisons of the orbital parameters
of individual merging events to be made.} we are confident that by
using halos containing ten or more particles we are obtaining an
accurate measure of the distributions.

\subsubsection{Radial search limits}

We also wish to test whether our imposed limits on the radial
separation of halos affects the distributions. To do this we use the
independent GIF and \gife\ $\tau$CDM $z=0$ outputs. Velocity
distributions are constructed from both simulation outputs using
radial search limits between $\Delta r=0.01$ and $\Delta r=0.20$ in
steps of $0.01$. We then compute the $\chi^2$ statistic comparing the
GIF simulation with $\Delta r=0.20$ to the \gife\ simulations with
$\Delta r < 0.2$, and vice versa. We find that the $\chi^2$ values
stay reasonably constant as the radial search limit is decreased, and
certainly show no signs of becoming significantly larger than
unity. As such, we conclude that the $\Delta r=0.2$ search limit is
sufficiently small to allow an accurate determination of the velocity
distributions.

\subsection{Trends}

Having established that the techniques employed in this paper are able
to accurately determine the distribution of orbital velocities for
infalling satellites we proceed to search for any dependence of those
distributions on the masses of the halos, redshift and cosmology. When
testing for such dependence we adopt the approach of varying only one
variable at a time, with the hope of isolating the cause of any trend
we discover. While this is crucial to developing an understanding of
the trends it significantly limits the number of comparisons that we
can make.

\subsubsection{Mass Dependence}

Since our distributions are constructed by combining the orbits of all
the halos, irrespective of mass, in a given simulation output it is
crucial that we first test for the presence of any trends with mass.
To test for mass-dependent trends we compare the GIF simulations with
the VIRGO and VLS simulations. These have identical cosmological
parameters, and we use halos with 10 or more particles in each
simulation. The only difference then is the particle mass and the
corresponding mass function of dark matter halos.

\begin{figure}
\psfig{file=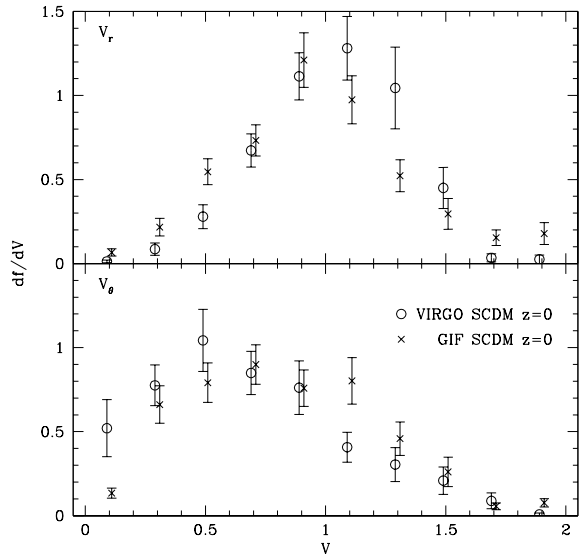,width=80mm}
\caption{Distributions of radial (upper panel) and tangential (lower
panel) velocities for the GIF and VIRGO SCDM $z=0$ outputs.}
\label{fig:compmass}
\end{figure}

We find evidence for mass-dependence in the distributions of orbital
parameters. Figure~\ref{fig:compmass} shows distributions of radial
and tangential velocities for GIF and VIRGO SCDM models at
$z=0$. There is a clear difference between the two, with the VIRGO
simulation showing larger radial and lower tangential velocities on
average. Unfortunately, our samples of mergers remain too small to
provide an accurate determination of the nature of the mass dependent
trends.

\subsubsection{Redshift and cosmology}

We next explore trends with redshift by comparing the results of
outputs from the same simulation at different epochs. Specifically we
compute $\chi^2$ for pairs of outputs which differ by at least 50\% in
$1+z$ to ensure that the samples are independent. We find strong
evidence for differences between these samples. However, as the mass
function of dark matter halos is a function of redshift, we cannot
disentangle any redshift-dependent trend from the known mass-dependent
trends. The current simulations do not possess enough halos to allow
us to select a sub-sample of mergers by mass at each redshift in order
to eliminate this problem. We also find significant differences
between models with different cosmological parameters, but again
cannot disentangle any possible mass-dependent trends. To fully
address these issues will require a set of custom N-body simulations
designed to allow us to explore changes in the orbital parameter
distributions in a controlled manner. (For example, the current
simulations have a variety of softening lengths, which may affect our
results. A dedicated set of N-body simulations could explore the
effects of this parameter on the distributions recovered.)

\subsubsection{Group Finding Algorithm}

We test for possible dependence on the group finding algorithm by
comparing distributions of orbital velocities from the VLS
$\Lambda$CDM simulation with halos found using the FOF group finder,
to those from the VIRGO $\Lambda$CDM simulation with halos found using
the SO group finder. We find no evidence for any systematic difference
between the distributions based upon these two group finders, and so
use the FOF algorithm throughout the remainder of this work.

\subsubsection{Linking Length}

\begin{figure}
\psfig{file=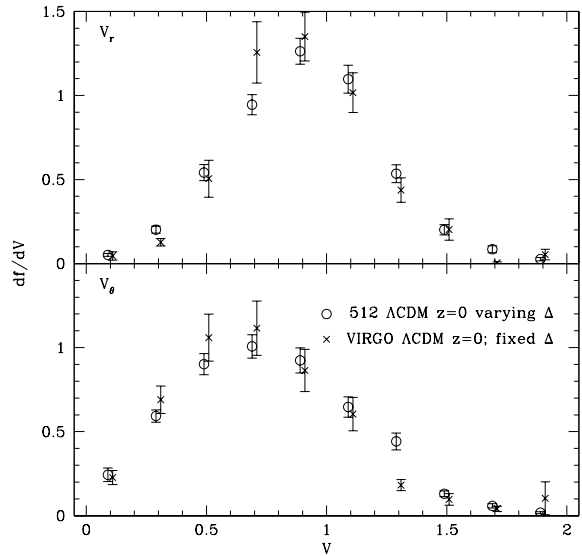,width=80mm}
\caption{Distributions of radial (upper panel) and tangential (lower
panel) velocities for the VIRGO OCDM and SCDM $z=0.10$ outputs.}
\label{fig:compvll}
\end{figure}

We test for possible dependence on the linking length by comparing
distributions of orbital velocities from the VLS $\Lambda$CDM
simulation with halos found using the fixed (varying) $\Delta$, to
those from the VIRGO $\Lambda$CDM simulation with halos found using
the varying (fixed) $\Delta$. The distributions are found to be
formally inconsistent with one another. Figure~\ref{fig:compvll} shows
a comparison. With the current statistical precision it is difficult
to determine the exact nature of the difference between fixed and
varying $\Delta$ distributions. We will thus not explore this further,
and will continue to use the varying $\Delta$ method.

\subsubsection{Halo Centring Algorithm}

\begin{figure}
\psfig{file=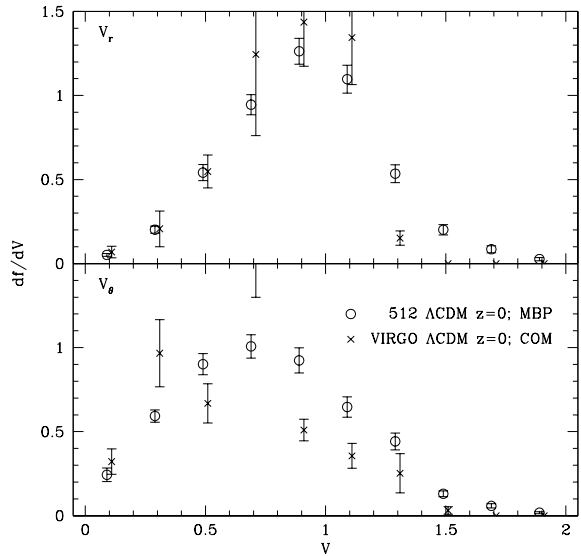,width=80mm}
\caption{Distributions of radial (upper panel) and tangential (lower
panel) velocities for the VLS and VIRGO $\Lambda$CDM $z=0$ outputs.}
\label{fig:compmb}
\end{figure}

We test for possible dependence on the halo centring algorithm by
comparing distributions of orbital velocities from the VLS and VIRGO
$\Lambda$CDM simulations with halos found using each algorithm (COM
and MBP). The distributions are again found to be formally
inconsistent with one another. In Figure~\ref{fig:compmb} we show a
comparison for the $z=0$ simulation outputs. The differences between
the two distributions are clearly visible. We find that the COM
algorithm typically produces distributions of radial and tangential
velocities which peak at lower values than the MBP algorithm. As we
demonstrated in Fig.~\ref{fig:halocentre}, the COM algorithm can
easily find an unrealistic origin in both position and velocity
space. Figure~\ref{fig:compmb} shows that this problem can
significantly affect the resulting distribution of orbital
parameters. We prefer to use the more robust MBP algorithm, and do so
throughout the remainder of this paper.

\subsection{Fitting Functions}
\label{sec:fitfunc}

The results presented in this work are potentially of great value to
any study involving the evolution of the substructure population of
cold dark matter halos. To facilitate their use in this way we provide
a simple fitting function which describes the two-dimensional
distribution of orbital velocities. Through simple variable
transformations this function also describes the distributions of
substructure energies, angular momenta, eccentricities etc.

We find that our measured two-dimensional distributions of orbital
velocities can be reasonably well fit with the following fitting
function:
\begin{equation}
f(v_{\rm r},v_\theta)=a_1 v_\theta \exp\left[ - a_2 (v_\theta-a_9)^2 -b_1(v_\theta) \{v_{\rm r}-b_2(v_\theta) \}^2 \right],
\label{eq:fitfunc}
\end{equation}
where
\begin{eqnarray}
b_1(v_\theta) & = & a_3 \exp\left[ - a_4 (v_\theta-a_5)^2 \right], \\
b_2(v_\theta) & = & a_6 \exp\left[ - a_7 (v_\theta-a_8)^2 \right]. \\
\end{eqnarray}
Note that this has a form similar to a two-dimensional
Maxwell-Boltzmann distribution for the tangential velocity and a
Gaussian for the radial velocity, as might be expected from the
results of \scite{vitvit02}. However, the mean and dispersion of the
radial velocity distribution are a function of the tangential
velocity, as is necessary to account for the correlation between these
two velocities found in our distributions.

\begin{figure}
\psfig{file=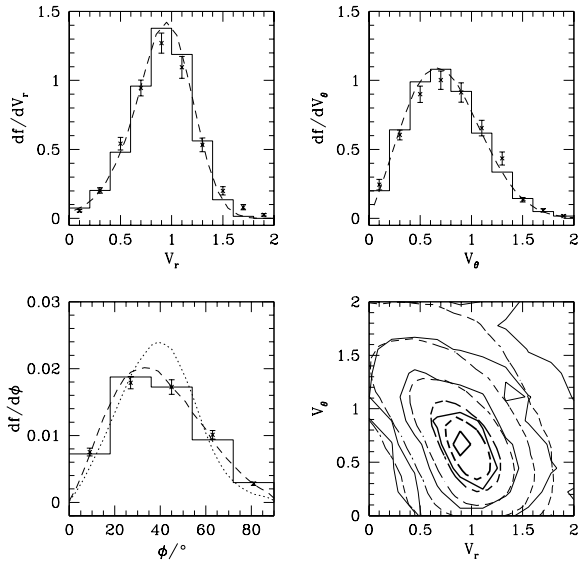,width=80mm}
\caption{Distributions of orbital parameters measured in the VLS plus
VIRGO $\Lambda$CDM $z=0$ outputs are shown by crosses. Upper left and
right-hand panels show distributions of radial and tangential
velocities respectively. The lower left-hand panel shows the
distribution of infall angles, while the lower right-hand panel shows
the two-dimensional distribution of radial and tangential velocities
(solid contours). Dashed lines show the fitting function, while
histograms show this function averaged over the same bins as used to
measure the distributions. The dotted line in the lower left-hand
panel indicates the distribution of infall angles that would occur if
correlations between $V_{\rm r}$ and $V_\theta$ were ignored.}
\label{fig:fit0}
\end{figure}

\begin{figure}
\psfig{file=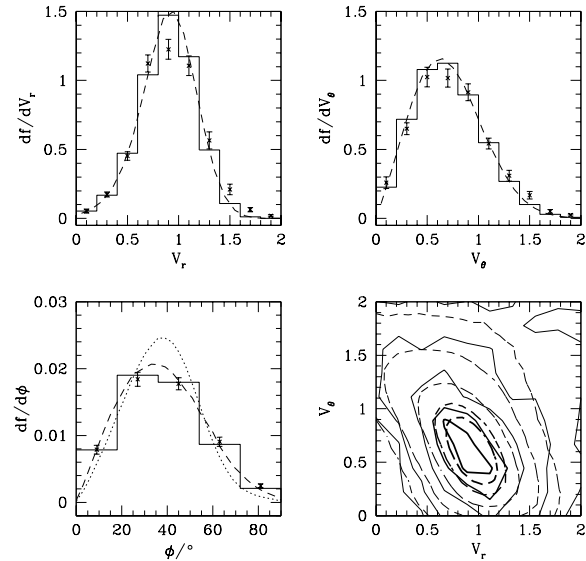,width=80mm}
\caption{As Figure~\protect\ref{fig:fit0} but for $z=0.5$.}
\label{fig:fit05}
\end{figure}

\begin{figure}
\psfig{file=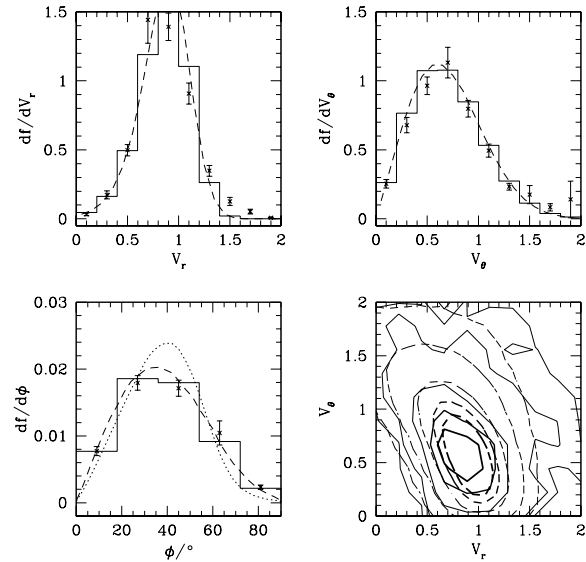,width=80mm}
\caption{As Figure~\protect\ref{fig:fit0} but for $z=1.0$.}
\label{fig:fit1}
\end{figure}

We have fit this function to distributions of orbits taken from the
combined VLS and VIRGO $\Lambda$CDM simulations (the VLS simulation is
the only one which provides sufficient signal to noise to make fitting
worthwhile). Figures~\ref{fig:fit0} through \ref{fig:fit1} show
distributions of orbital velocities together with the fitting
function, while Table~\ref{tb:fitpar} lists the parameter values used
in the fits.

\begin{table}
\caption{Parameters of the fitting function given in
eqn.~(\protect\ref{eq:fitfunc}). Each column lists parameters which
best fit distribution of orbital parameters in the combined VLS and
VIRGO $\Lambda$CDM simulations at the specified redshift.}
\label{tb:fitpar}
\begin{center}
\begin{tabular}{cr@{.}lr@{.}lr@{.}l}
\hline
 & \multicolumn{6}{c}{Redshift} \\
Parameter & \multicolumn{2}{c}{$0.0$} & \multicolumn{2}{c}{$0.5$} & \multicolumn{2}{c}{$1.0$} \\
\hline
$a_1$ & 3&90  & 4&46   & 6&38    \\
$a_2$ & 2&49  & 2&98   & 2&30    \\
$a_3$ & 10&2  & 11&0   & 18&8    \\
$a_4$ & 0&684 & 1&11   & 0&506   \\
$a_5$ & 0&354 & 0&494  & -0&0934 \\
$a_6$ & 1&08  & 1&16   & 1&05    \\
$a_7$ & 0&510 & 0&261  & 0&267   \\
$a_8$ & 0&206 & -0&279 & -0&154  \\
$a_9$ & 0&315 & 0&331  & 0&157   \\
\hline
\end{tabular}
\end{center}
\end{table}

\subsection{Other quantities}

Other quantities which characterise the satellite orbits are easily
derived from the two velocities $V_{\rm r}$ and $V_\theta$. For
convenience, we list below expressions for several other orbital
parameters in terms of these velocities.\\

\noindent \textbf{Specific energy:}
\begin{equation}
E = -1+{1\over 2 f_2} \left( V_{\rm r}^2 +{(2-f_2)^2 \over f_2^2} V_{\theta}^2 \right),
\end{equation}

\noindent \textbf{Specific angular momentum:}
\begin{equation}
J = V_{\rm \theta}
\end{equation}

\noindent \textbf{Eccentricity:}
\begin{equation}
e = {V_\theta^2\over f_2} \sqrt{\left( 1 - {f_2 \over V_\theta^2} \right)^2 + \left( {V_{\rm r} \over V_\theta}\right)^2}
\label{eq:fecc}
\end{equation}

\noindent \textbf{Circularity:}
\begin{equation}
\epsilon = V_\theta \sqrt{{2f_2-V_{\rm r}^2-V_\theta^2 \over 2f_2-1}}
\label{eq:fcirc}
\end{equation}

\noindent \textbf{Semi-major axis:}
\begin{equation}
a = {f_2 \over 2 f_2 - V_{\rm r}^2 - V_\theta^2}
\end{equation}

\noindent \textbf{Pericentric distance:}
\begin{equation}
r_{\rm peri} = \left[ {f_2\over V_\theta^2} + \sqrt{ \left(1 - {f_2 \over V_\theta^2} \right)^2 +  \left( {V_{\rm r}\over V_{\rm t}}\right)^2 }  \right]^{-1}
\end{equation}

\noindent \textbf{Apocentric distance:}
\begin{equation}
r_{\rm apo} = \left[ {f_2\over V_\theta^2} - \sqrt{ \left(1 - {f_2 \over V_\theta^2} \right)^2 +  \left( {V_{\rm r}\over V_{\rm t}}\right)^2 }  \right]^{-1}
\end{equation}

\subsubsection{Eccentricity and semi-major axis}

\begin{figure*}
\begin{tabular}{cc}
\psfig{file=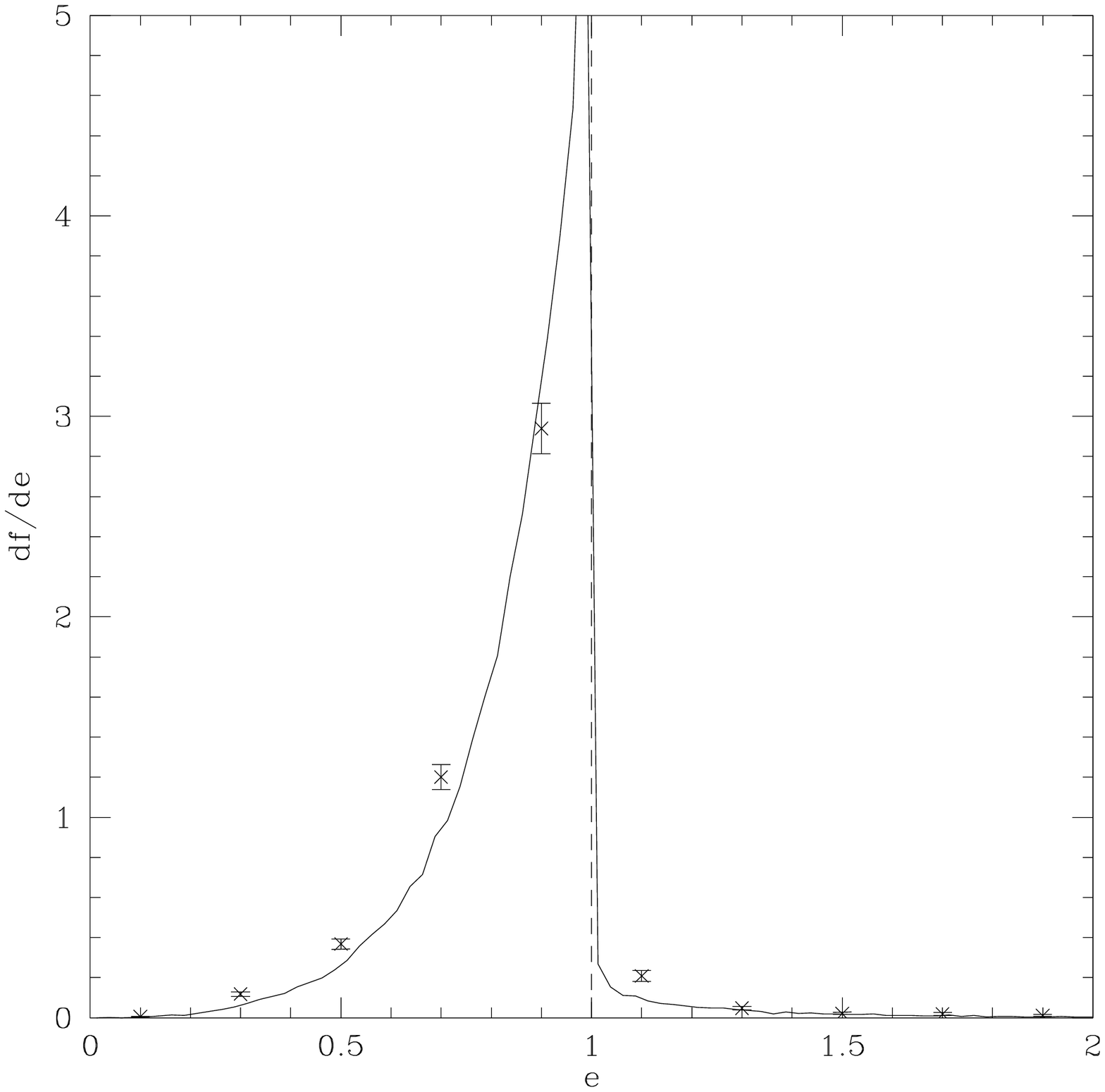,width=80mm} & \psfig{file=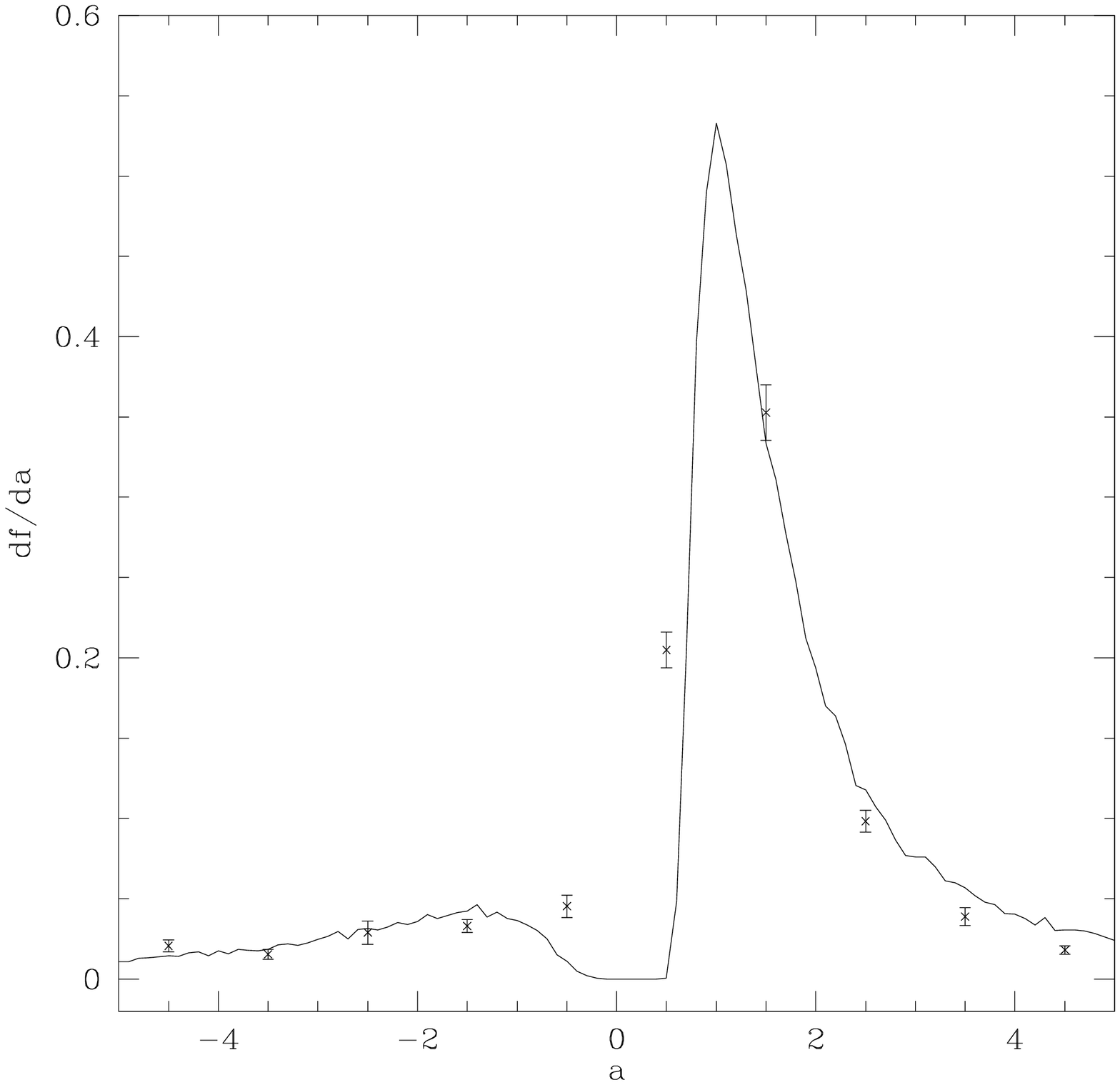,width=80mm}
\end{tabular}
\caption{Distributions of eccentricity (left-hand panel) and
semi-major axis (right-hand panel) for the VLS plus VIRGO $\Lambda$CDM
$z=0$ outputs are shown by the crosses with errorbars. The solid lines
indicate the distribution resulting from the fitting formula of
eqn.~\protect\ref{eq:fitfunc}. The vertical dashed line the left-hand
panel indicates parabolic orbits, and so the division between bound
($\epsilon < 1$) and unbound ($\epsilon > 1$) orbits. In the right
hand panel, negative values of $a$ correspond to unbound orbits. In
this case the semi-major axis of the hyperbolic orbit is $|a|$.}
\label{fig:eccsm}
\end{figure*}

We have presented results for radial and tangential velocities, but of
course can just as easily examine invariant parameters of the orbits,
such as eccentricity and semi-major axis. Figure~\ref{fig:eccsm} shows
distributions of these two parameters from the VLS $\Lambda$CDM $z=0$
output, together with the distributions implied by our fitting
function. Our distribution of eccentricities is qualitatively, but not
quantitatively, in agreement with that presented in the first version
of the preprint (i.e. {\tt astro-ph/0309611} version 1, hereafter
Khochfar \& Burkert (2004) v.1) by \scite{khochfar04}, with most
orbits being close to parabolic ($e=1$). We find that almost half of
all orbits have $e=1\pm 0.1$, a somewhat smaller fraction than the
70\% given by \scite{khochfar04} v.1.

\begin{figure}
\psfig{file=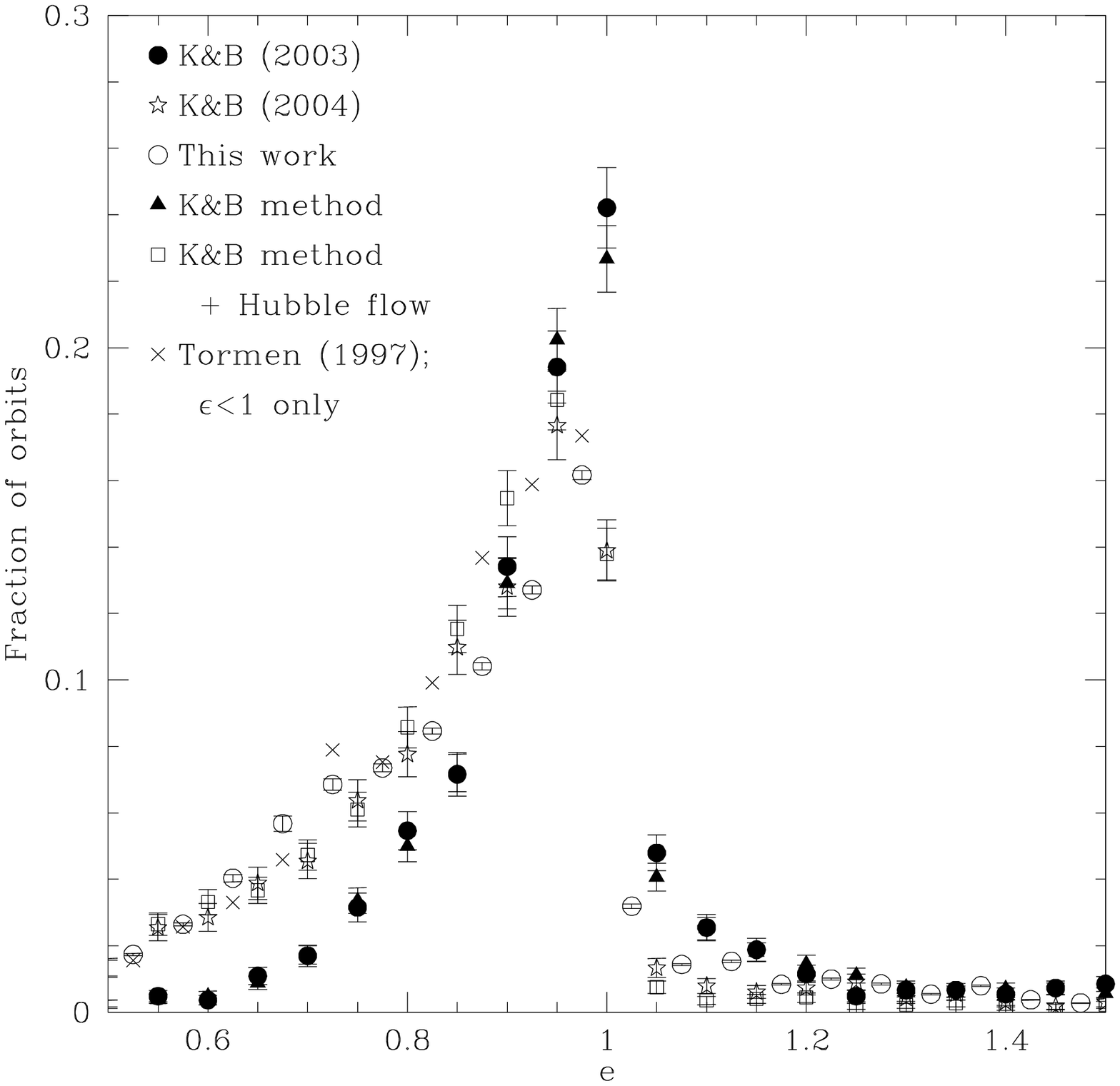,width=80mm}
\caption{The distribution of orbital eccentricities. The quantity
shown is the fraction of orbits in each eccentricity bin
(i.e. following the format of Figure~1 of
\protect\pcite{khochfar04}). Filled circles indicate the results of
\protect\scite{khochfar04} v.1, while crosses show the results of
\protect\scite{tormen97}. Open circles are results from this work
combining all redshifts from the GIF $\Lambda$CDM simulation using the
MBP halo centring algorithm. Filled triangles show our implementation
of Khochfar \& Burkert's (2004) methods when no Hubble flow is added
to the velocities of particles in the N-body simulations, while open
squares show the same with the Hubble flow added. Stars indicate the
results of \protect\scite{khochfar04} which represent the same
calculation as \protect\scite{khochfar04} v.1 revised to include the
Hubble flow.}
\label{fig:ecccomp}
\end{figure}

In fact, as we show in Fig.~\ref{fig:ecccomp} our results are
significantly different from those of \scite{khochfar04}
v.1. Comparing results from this work with those of \scite{khochfar04}
v.1 we find that our results, though peaked around $e=1$, are more
broadly distributed. \scite{khochfar04} use a different approach to
finding merging halos than we do\footnote{Briefly, they locate the
progenitors of a given halo at a slightly earlier redshift. They then
measure the orbital properties of these progenitors, providing they
are separated by more than the sum of their virial radii. To ensure
that the apparently merging halos are not merely undergoing an unbound
``fly-by'' they also check that the centres of the halos have not
moved further apart by a later redshift.} and this could potentially
influence the results obtained. We have implemented Khochfar \&
Burkert's (2004) methods on the GIF $\Lambda$CDM simulations to test
for any systematic effects caused by the difference in methods. We
have checked that our implementation produces eccentricities identical
to theirs (Khochfar, private communication). \scite{khochfar04} v.1
did not add on the Hubble flow velocity to the motions of halos
(Khochfar, private communication). Using the \scite{khochfar04}
methods we obtained the distributions shown by filled triangles and
open squares in Fig.~\ref{fig:ecccomp}. (Filled triangles have no
Hubble flow added to halo motions, while open squares do have the
Hubble flow added.) We find that we are able to reproduce the results
of Khochfar \& Burkert when using only halo peculiar velocities in our
calculations, and are able to reproduce our own results when the
Hubble flow is included.

As a second check, we have taken the distribution of orbital
circularities found by \scite{tormen97}, who used techniques similar
to \scite{khochfar04}, and converted these into eccentricities using
eqns.~(\ref{eq:fecc}) and (\ref{eq:fcirc}) (assuming
$f_2=1$). Correcting for the fact that orbits with $e>1$ are not
included in the distribution of \scite{tormen97} we find an
eccentricity distribution as shown by the crosses in
Fig.~\ref{fig:ecccomp}.

We conclude that these two different approaches to determining
distributions of halo orbital parameters produce consistent results,
providing they attempt to measure the same quantities. The differences
between the distributions of eccentricities reported here and by
\scite{khochfar04} v.1 can be traced to the choice of whether to
include the Hubble flow in particle velocities (as we did), or to use
peculiar velocities, as did \scite{khochfar04} v.1\footnote{As a
result of discussions regarding these differences, Khochfar \& Burkert
have revised their calculations to include the Hubble flow (see the
published \protect\pcite{khochfar04} or version 2 of the
preprint). Their results are then in good agreement with those found
in this work, as shown by the stars in Fig.~\ref{fig:ecccomp}.}.

\subsubsection{Correlations between pairs of infalls}

\begin{figure}
\psfig{file=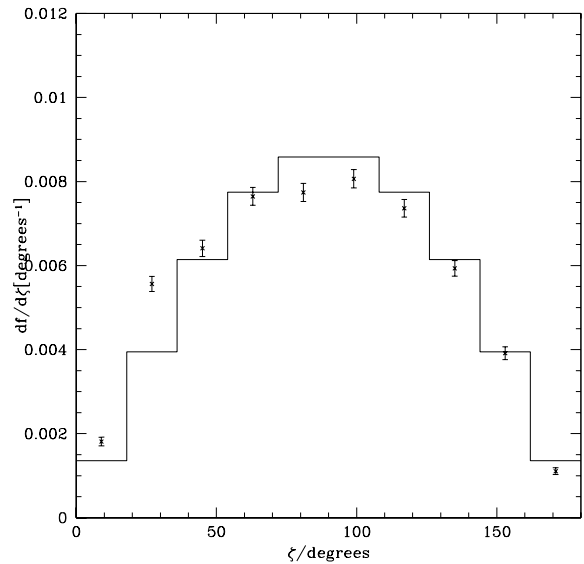,width=80mm}
\caption{The distribution of angles between the infall directions of
pairs of satellites merging onto the same host halo. Points show the
distribution measured by summing results from all simulation outputs
while the histogram indicates the expectation for uncorrelated infall
directions.}
\label{fig:pair}
\end{figure}

We can test for correlations between the infall directions of pairs of
satellites merging into the same halo. Figure~\ref{fig:pair} shows the
distribution of angles $\phi$ between the radius vectors of pairs of
satellites merging into the same host halo.\footnote{In this and
subsequent figures exploring angles between pairs of satellites or
satellites and the host halo spin we do not include our usual weights
when constructing the distributions. Our weights reflect the fact
that, due to the snapshot sampling provided by the N-body simulations)
we do not see all mergers, but only those which occur within a short
time after the snapshot. When constructing velocity (or eccentricity,
semi-major axis etc.) distributions, the weighting used corrects for
the unobserved population of mergers. To make the same correction when
considering the infall angles here we must supplement the weight with
an assumption about the angular distribution of the unobserved
mergers. We make the simple assumption that the unobserved mergers
have the same angular distribution as those which we do observe. As
such, the resulting angular distribution is found from the observed
mergers without any weights. Note that this assumption may be
incorrect---for example, if the angular distribution correlates with
infall velocity---but is at least simple.} Note that we have summed
the results from all simulation outputs to obtain this
distribution. This is permissible as our aim here is to search for any
deviation from uncorrelated infall directions. As such, it does not
matter if the different outputs are correlated in different ways---we
would still see a difference from the null hypothesis of no
correlations. The distribution appears to differ significantly from
that expected if there were no correlations between infall
directions. This correlation between infall directions is
qualitatively as expected if mergers tend to occur along filaments,
i.e. there is an enhancement in the number of mergers at small angles,
$\zeta \lsim\, 30^\circ$, with a corresponding suppression of mergers
with angles around $90^\circ$.

\subsubsection{Spin alignments}

\begin{figure*}
\begin{tabular}{cc}
\psfig{file=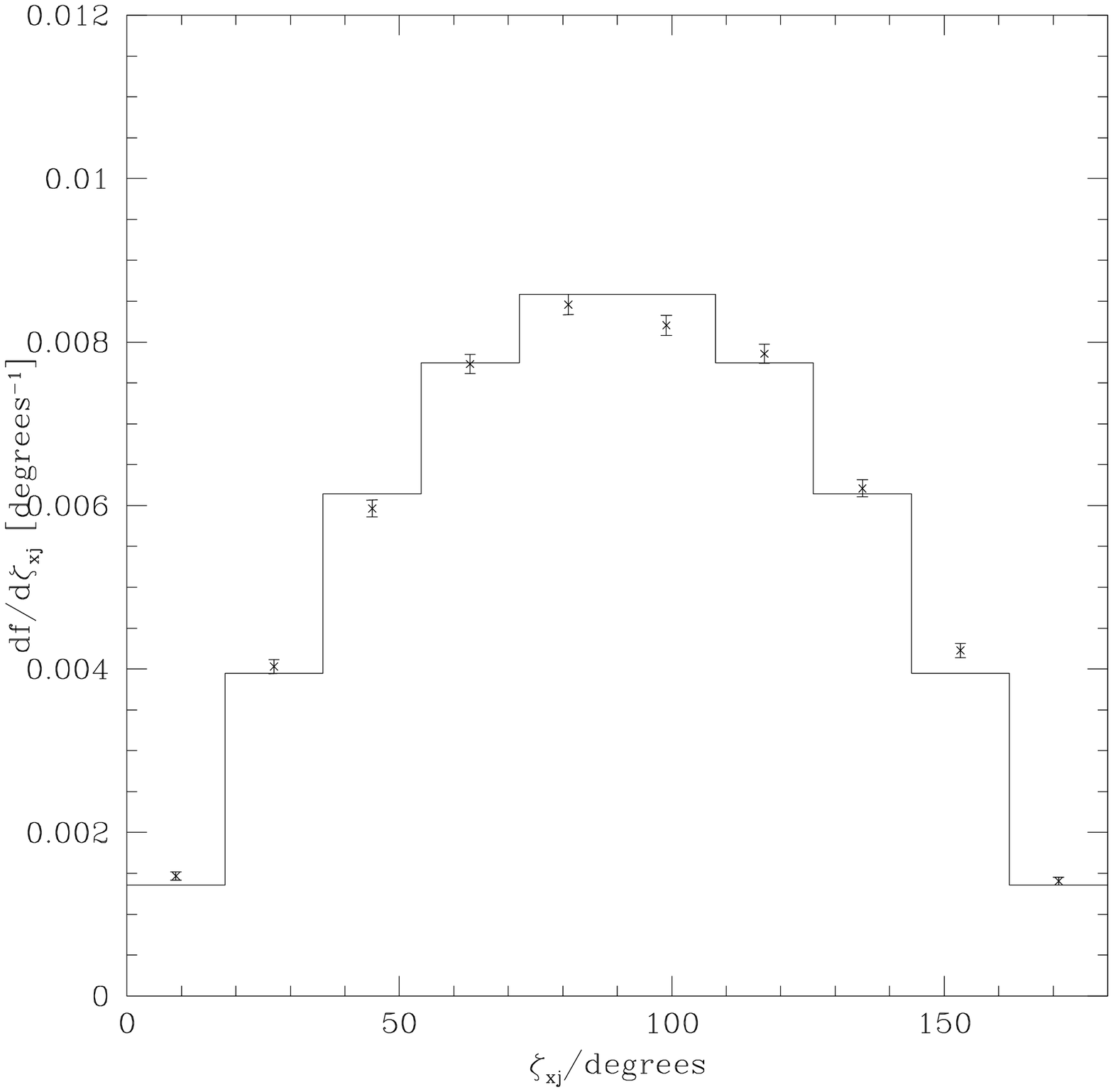,width=80mm} & \psfig{file=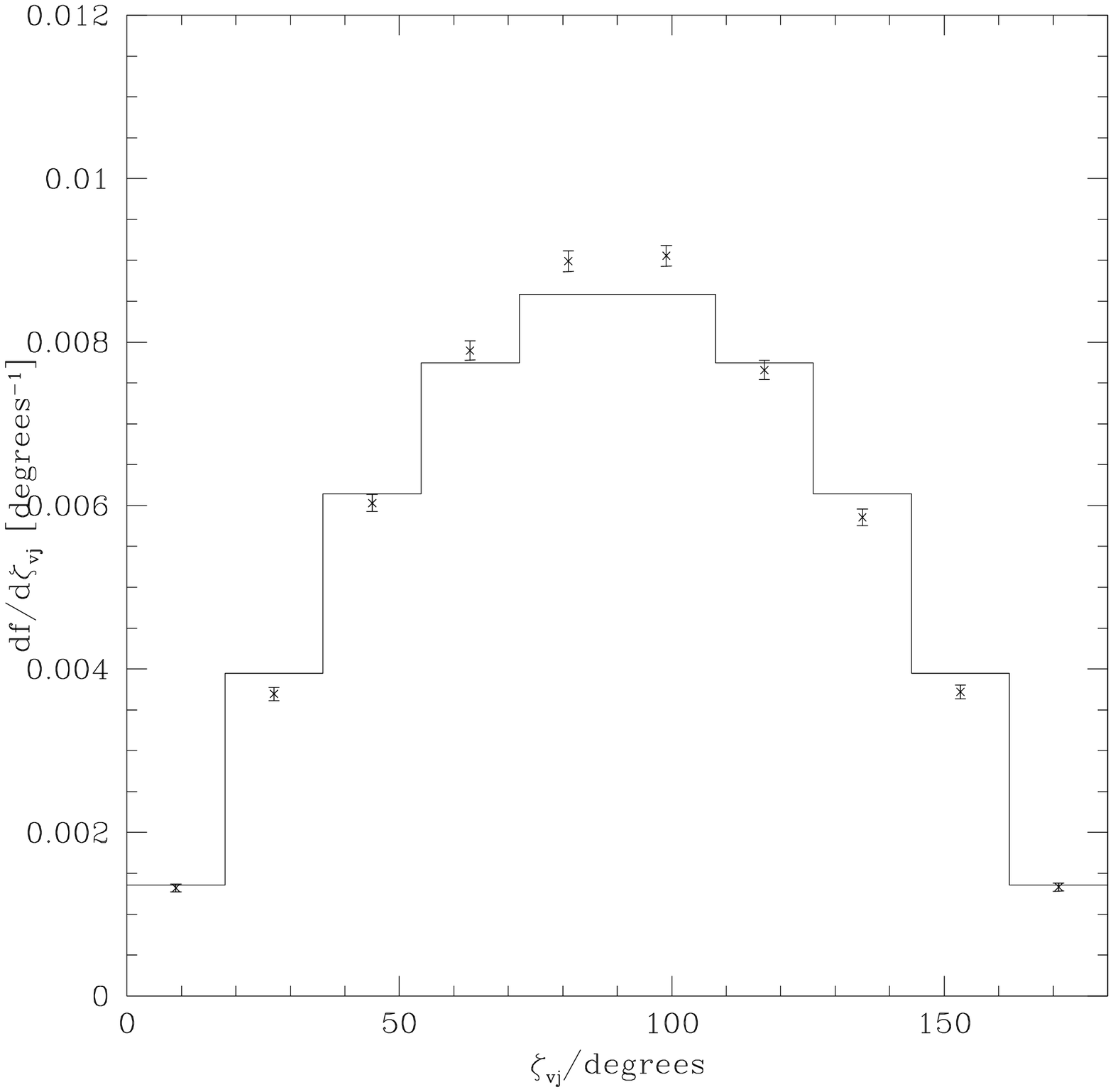,width=80mm}
\end{tabular}
\caption{The distribution of angles between the infall direction
(left-hand panel) and infall velocity (right-hand panel) of satellites
and the angular momentum of the host halo. Points show results
measured by summing merger events from all simulation outputs while
histograms show the expectation when no correlations are present.}
\label{fig:spin}
\end{figure*}

Finally, we can examine correlations between the infall direction and
velocity of satellites and the spin angular momentum vector of the
host halo. Figure~\ref{fig:spin} shows the resulting distributions. We
find marginal evidence for deviations from a uniform distribution on
the sphere. In particular, there is a suggestion that merging
satellites have a tendency to have velocities normal to the spin axis
of their host halo.

To assess the validity of these results will require a better
calibration of our errors. For example, the direction of the spin
vector may be poorly determined for low mass halos, a contribution to
the errors that we do not take into account. (Although this effect
should presumably weaken any correlations, implying that the true
correlations are stronger than those we measure.)

\section{Discussion}
\label{sec:discuss}

We have described methods for determining the orbital parameters of
dark matter halos at the point of merging with a larger
system. Previous studies of the orbital properties of merging halos
have typically considered the orbits after merging with the host halo,
in which case the orbits will have changed due to dynamical
friction. Other studies \cite{tormen97,khochfar04} used techniques
which are restricted to simulations with closely spaced outputs if
they are to be accurate. Furthermore, we have analysed a substantially
larger number of orbits than has been previously possible to obtain
improved statistical precision. This allows us to characterise in
detail the two-dimensional distribution of infall velocities.

Our analysis pays particular attention to carefully identifying halos
and their centres. We find that is is important to accurately identify
the centre of the halo in both position and velocity space, and adopt
a similar minimum ``energy'' definition for both of these. We have
demonstrated that our results are unbiased by effects of particle
number or radial search limit. In this work, we have focused on the
two-dimensional distribution of radial and tangential velocities which
we show has a relatively simple form. A fitting formula that describes
this distribution is presented and should prove immensely valuable in
future studies of satellite orbits.

Our methods could be improved upon in several ways. A set of
simulations run with measurements of orbital parameters in mind would
allow a better determination of the accuracy of our error
estimates. More and larger simulations would also improve the
statistical accuracy of our measurements and permit us to quantify the
trends with, for example, mass that are apparent in the
distributions. Finally, a more detailed treatment of the evolution of
the satellite orbits (including the effects of an extended,
non-spherical host halo and dynamical friction) would remove sources
of systematic error in our measurements. All of these factors will be
the subject of a future paper.

We have presented evidence for the presence of trends with mass (and,
perhaps, with redshift and cosmological parameters) in this
distribution, although we are currently unable to accurately
characterize these trends. Larger samples of N-body halo mergers will
allow us to both characterise these mass trends and to select
sub-samples with a narrow range in mass to permit trends with redshift
and cosmological parameters to be examined.

We have also explored the distribution of eccentricities and
semi-major axes. We find that the eccentricity distribution is peaked
around parabolic orbits ($e=1$). This is qualitatively in agreement
with the work of \scite{khochfar04} v.1. However, we find quantitative
disagreements with their distribution of eccentricities. This
disagreement has been traced to the fact that the Hubble flow was
included in our calculations, while it was not included in those of
\scite{khochfar04} v.1. Once Hubble flow is included, as in the final
version of \scite{khochfar04}, the results of the two studies are in
excellent agreement. Our distributions of eccentricities and
tangential velocities are also in good agreement with those from
\scite{tormen97} and \scite{vitvit02} respectively.

Finally, we searched for correlations between the infall directions of
pairs of satellites and between the infall positions and velocities of
satellites and the angular momentum of their host halo. We find
evidence that satellites infalling onto a given host tend to arrive
from similar directions, compatible with the hypothesis that (at least
some) infall occurs along filaments. We find marginal evidence that
infall directions and direction of motion are aligned with the
spin-axis of the host halo, although a more thorough study would be
required to both confirm and interpret this possible correlation.

The evolution of sub-structures in cold dark matter halos is currently
a topic of great interest. The tools provided in this work should
prove of great value in further such studies while the techniques
described should permit more accurate estimates of orbital parameter
distributions (including dependences on halo mass, spin, redshift,
cosmology etc.) to be constructed.

\section*{Acknowledgements}

AJB acknowledges valuable discussions with Carlos Frenk, Joel Primack
and T.~J.~Cox and thanks Sadegh Khochfar for providing data for
Figure~\ref{fig:ecccomp} in electronic form and for extensive
discussions regarding eccentricity distributions. AJB also
acknowledges support from a Royal Society University Research
Fellowship. The simulations in this paper were carried out by the
Virgo Supercomputing Consortium using computers based at Computing
Centre of the Max-Planck Society in Garching and at the Edinburgh
Parallel Computing Centre. The data are publicly available at {\tt
www.mpa-garching.mpg.de/NumCos}.


\begin{thebibliography}{}
\bibitem[Abadi, Bower \& Navarro <2000>]{abadi00}Abadi~M.~G., Bower~R.~G., Navarro~J.~F., 2000, MNRAS, 314, 759
\bibitem[Benson et al. <2001>]{benson01}Benson~A.~J., Frenk~C.~S., Baugh~C.~M., Cole~S., Lacey~C.~G., 2001, MNRAS, 327, 1041
\bibitem[Benson et al. <2002>]{benson02}Benson~A.~J., Lacey~C.~G., Baugh~C.~M., Cole~S., Frenk~C.~S., 2002, MNRAS, 333, 156
\bibitem[Benson et al. <2004>]{benson04}Benson~A.~J., Lacey~C.~G., Frenk~C.~S., Baugh~C.~M., Cole~S., 2004, MNRAS, 351, 1215
\bibitem[Chiba <2002>]{chiba01}Chiba~M., 2002, ApJ, 565, 17
\bibitem[Dalal \& Kochanek <2002>]{dal01}Dalal~N., Kochanek~C.~S., 2002, ApJ, 572, 25 
\bibitem[Davis et al. <1985>]{defw}Davis~M., Efstathiou~G., Frenk~C.~S., White~S.~D.~M., 1985, ApJ, 293, 371
\bibitem[Diemand, Moore \& Stadel <2004>]{diemand04}Diemand~J., Moore~B., Stadel~J., 2004, MNRAS in press (astro-ph/0402160)
\bibitem[Eke, Cole \& Frenk <1996>]{eke96}Eke~V.~R., Cole~S., Frenk~C.~S., 1996, MNRAS, 282, 263
\bibitem[Font et al. <2001>]{font01}Font~A.~S., Navarro~J.~F., Stadel~J., Quinn~T., 2001, ApJ, 563, L1
\bibitem[Gao et al. <2004>]{gao04}Gao~L., White~S.~D.~M., Jenkins~A., Stoehr~F., Springel~V., 2004, submitted to MNRAS (astro-ph/0404589)
\bibitem[Ghigna et al. <1998>]{ghigna98}Ghigna~S., Moore~B., Governato~F., Lake~G., Quinn~T., Stadel~J., 1998, MNRAS, 300, 146
\bibitem[Jenkins et al. <1998>]{jenkins98}Jenkins~A., Frenk~C.~S., Pearce~F.~R., Thomas~P.~A., Colberg~J.~M., White~S.~D.~M., Couchman~H.~M.~P., Peacock~J.~A., Efstathiou~G., Nelson~A.~H., 1998, ApJ, 499, 20
\bibitem[Jenkins et al. <2001>]{jenkins01}Jenkins~A., Frenk~C.~S., White~S.~D.~M., Colberg~J.~M., Cole~S., Evrard~A.~E., Couchman~H.~M.~P., Yoshida~N., 2001, MNRAS, 321, 372
\bibitem[Kauffmann et al. <1999.>]{kauffmann99}Kauffmann~G., Colberg~J.~M., Diaferio~A., White~S.~D.~M., 1999, MNRAS, 303, 188
\bibitem[Khochfar \& Burkert <2004>]{khochfar04}Khochfar~S., Burkert~A., 2004, MNRAS submitted (astro-ph/0309611)
\bibitem[Klypin et al. <1998>]{klypin98}Klypin~A., Gottl\"ober~S., Kravtsov~A.~V., Khokhlov~A.~M., ApJ, 516, 530
\bibitem[Lacey \& Cole <1993>]{lc93}Lacey~C.~G., Cole~S., 1993, MNRAS, 262, 627
\bibitem[Lacey \& Cole <1994>]{lc94}Lacey~C.~G., Cole~S., 1994, MNRAS, 271, 676
\bibitem[Metcalf \& Madau <2001>]{metmad01}Metcalf~R.~B., Madau~P., 2001, ApJ, 563, 9
\bibitem[Moore, Lake \& Katz <1998>]{moore98}Moore~B., Lake~G., Katz~N., 1998, ApJ, 495, 139
\bibitem[Moore et al. <1999>]{moore99}Moore~B., Ghigna~S., Governato~F., Lake~G., Quinn~T., Stadel~J., Tozzi~P., 1999, ApJ, 524, 19
\bibitem[Navarro, Frenk \& White <1997>]{NFW}Navarro~J.~F., Frenk~C.~S., White~S.~D.~M., 1997, ApJ, 490, 493
\bibitem[Peebles <1980>]{peebles80}Peebles~P.~J.~E., 1980, The Large Scale Structure of the Universe, Princeton Univ. Press, Princeton, NJ
\bibitem[Taylor \& Babul <2004>]{taybab04}Taylor~J.~E., Babul~A., 2001, MNRAS, 348, 811
\bibitem[Tormen <1997>]{tormen97}Tormen~G., 1997, MNRAS, 290, 411
\bibitem[Tormen, Diaferio \& Syer <1998>]{tormen98}Tormen~G., Diaferio~A., Syer~D., 1998, MNRAS, 299, 728
\bibitem[van den Bosch et al. <1999>]{vdb99}van den Bosch~F.~C., Lewis~G.~F., Lake~G., Stadel~J., 1999, ApJ, 515, 50
\bibitem[Vitvitska et al. <2002>]{vitvit02}Vitvitska~M., Klypin~A.~A., Kravtsov~A.~V., Wechsler~R.~H., Primack~J.~R., Bullock~J.~S., 2002, ApJ, 581, 799
\bibitem[Zhang et al. <2002>]{zhang02}Zhang~B., Wyse~R.~F., Stiavelli~M., Silk~J., 2002, MNRAS, 332, 647
\end{thebibliography}
\end{document}